\documentclass[12pt]{article}
\usepackage{fullpage}
\usepackage{graphicx}
\usepackage{hyperref}
\usepackage{natbib}
\newif\ifagu
\def\gesim{\ \hbox to 0 pt{\raise .6ex\hbox{$>$}\hss}\lower.5ex\hbox{$\sim$}\ }
\def\lesim{\ \hbox to 0 pt{\raise .6ex\hbox{$<$}\hss}\lower.5ex\hbox{$\sim$}\ }
\begin{document}
\title{Electron velocity distribution instability in magnetized plasma
  wakes
  and artificial electron mass}
\author{I H Hutchinson\\
Plasma Science and Fusion Center and\\
Department of Nuclear Science and Engineering,\\
 Massachusetts Institute of Technology, Cambridge, MA 02139, USA}
\date{}
\maketitle

\begin{abstract}
The wake behind a large object (such as the moon) moving rapidly
through a plasma (such as the solar wind) contains a region of
depleted density, into which the plasma expands along the magnetic
field, transverse to the flow. It is shown here that (in addition to
any ion instability) a bump-on-tail which is unstable appears on the
\emph{electrons}' parallel velocity distribution function because of the convective
non-conservation of parallel energy. It arises regardless of any
non-thermal features on the external electron velocity
distribution. The detailed electron distribution function throughout
the wake is calculated by integration along orbits; and the substantial
energy level of resulting electron plasma (Langmuir) turbulence is
evaluated quasilinearly. It peaks near the wake axis.  If the mass of
the electrons is artificially enhanced, for example in order to make
numerical simulation feasible, then much more unstable electron
distributions arise; but these are caused by the unphysical mass
ratio.
\end{abstract}

\ifagu
\begin{article}
\fi

\section{Introduction}

Magnetized plasma wakes have attracted renewed interest recently
because of new measurements of the solar wind in the vicinity of the
moon \cite{Halekas2011,Wiehle2011}, but also because of their wider applications to space-craft,
dust grains, and laboratory flow measurement probes \cite{Patacchini2010,Hutchinson2010}. This paper
explores the effects of supersonic wakes on electron parallel-velocity
distributions and the instability that is induced.

We consider an insulating object whose size, $R$, is much greater than
the Debye length, $\lambda_{De}=\sqrt{\epsilon_0T_e/e^2 n_e}$, so that
with the exception of a negligible thickness sheath, the surrounding
region can be considered quasi-neutral. We suppose that the object is
moving through a magnetized plasma in which the ion Larmor radius is
also much smaller than $R$. The dynamics parallel to the magnetic
field can then be separated from the perpendicular for the ions and
even more definitively for the electrons (whose Larmor radius is even
smaller). This situation is very representative, for example, of the
moon and other unmagnetized planetary bodies moving through the solar
wind. The moon's radius is 1730km; the Debye length is of order 10m
and the ion Larmor radius of order
40km\cite{Ogilvie1996,Halekas2005}. Although the solar wind has ratio
of plasma to magnetic pressure, $\beta\sim 1$, and thus the wake
experiences significant magnetic perturbations (of order 10\% at 4
moon radii
\cite{Wiehle2011}), these will be ignored. The magnetic field here is
taken to be uniform and simply one-dimensionalizes the problem. Many
other Alfv\'enic phenomena must be accounted for if non-zero beta
effects are to be incorporated (see
e.g. \cite{Kallio2005,Travnicek2005}), but we here focus on the
electrostatic phenomena, which are an important part of the picture.

We will assume for simplicity during the development that the
direction of object motion, or equivalently, in the frame of the
object, the plasma drift, is at right angles to the magnetic field. It
is shown in section \ref{appendix}, how the results we obtain can
immediately, rigorously, be generalized and applied to oblique field-drift
alignment, which is more typical of the solar wind.

In the rest-frame of the object, the electrons and ions sweep rapidly past
under the influence of a drift electric field perpendicular to both
the magnetic field and plasma drift velocity $v_\perp$. Typical solar wind
velocity is of order 400km/s, roughly ten times the ion sound
speed.
\begin{figure}[htp]
  \hbox to \hsize{\hss
  \includegraphics[width=3.2in]{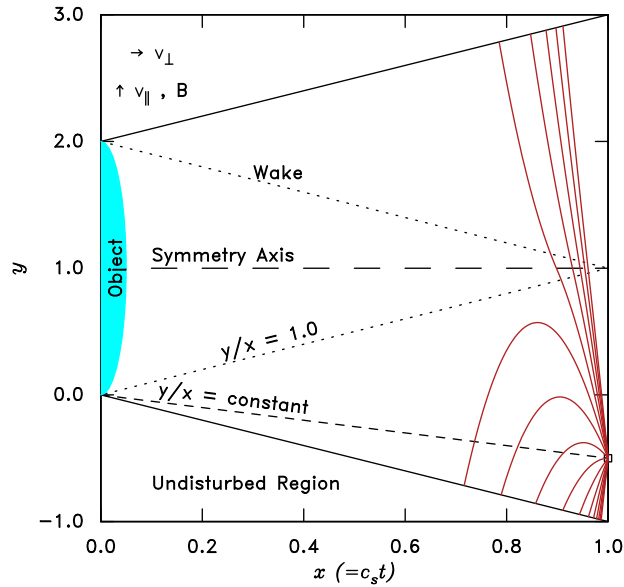}
  \hss}
  \caption{Geometry of a wake, illustrating electron orbits (curves) tracked
    back from position (1,0.5), and self similar lines ($y/x=const$)
    on which the potential is constant. \label{geom}}
  \end{figure}
 Fig.\ \ref{geom} illustrates the geometry of a wake in which
the spatial coordinate in the drift direction ($x$) has been divided
by $v_\perp/c_s$, where $c_s$ ($=\sqrt{T_e/m_i}$ for a Maxwellian) is
the (cold ion) sound speed. Thus $dx$ is equal to $c_sdt$ when moving at the
constant speed $v_\perp$. In these units the object is foreshortened
by a ratio equal to the perpendicular Mach number and its wake $x$-extent
is approximately equal to its radius.
We will assume the
drift Mach number is large enough to justify ignoring the object's
radius of curvature at its edges. This is the only place where the
treatment is limited to supersonic cases. Subsonic flow gives rise to
object \emph{elongation} in this scaled coordinate system, and its curvature
cannot then safely be ignored (unless it starts as a flat disk rather
than a sphere). Note, though, that the mechanisms we explore are still
active in subsonic cases, even though our quantitative treatment cannot be
expected to be accurate.

The physics close to the edge of the object, before particle streams
from above and below have begun to overlap, is well represented as the
expansion of a plasma into vacuum, which has been well understood for
a long time \cite{Gurevich1969,Gurevich1975,Samir1983}. It has also
been shown more recently \cite{Hutchinson2008b} that the additional
drifts arising from self-consistent electric field in the magnetized
case can be ignored, reducing the problem to two dimensions. Using
quasineutrality, $n_e=n_i$, the ion dynamics can be solved
self-consistently, analytically by ignoring the ion pressure or
numerically in one dimension with full ion kinetics
\cite{Gurevich1969,Patacchini2009}, and taking the potential to be
given by a direct relationship with the electron density such as the
Boltzmann relation $n_e \propto \exp(e\phi/T_e)$, or more generally a
polytropic assumption \cite{Sack1987,Manfredi1993} $p_e\propto
n_e^\gamma$.

To justify these simple electron models one invokes (1) Liouville's
theorem that, if collisionless, the electron distribution function is
constant on an orbit, and (2) the presumption that all (or nearly all)
the electron orbits can be tracked back to the undisturbed plasma. One
must also invoke (3) an electron parallel energy conservation
equation, normally in the form ${1\over2}m_e v_\parallel^2 -e \phi =
constant$, where $m_e$ is the electron mass, $v_\parallel$ its parallel
velocity, $-e$ its charge, and $\phi$ the electric potential. When the
external undisturbed parallel distribution function far from the
object (where $\phi=0$), is a function only of $v_\parallel^2$, then
with this conservation law, the distribution at a position where the
potential $\phi$ is non-zero is also of this form, but appropriately
shifted in $v_\parallel^2$. The result is that if the distribution
starts Maxwellian it remains Maxwellian,; and similarly that if it
starts having a so-called Kappa distribution \cite{Hellberg2009}
\begin{equation}
  f(v) \propto (1+v_\parallel^2/\kappa\theta^2)^{\kappa+1} ,
\end{equation}
(which can be considered a generalized Lorentz distribution) then it
remains a Kappa distribution with the same $\kappa$ (but varying
$\theta$) \cite{Meyer-Vernet1995}. A Kappa distribution gives rise to polytropic density
variation with $\gamma=1-2/(2\kappa-1)$. In any case the density is
a well-defined function of potential, not of position explicitly.

The purpose of this paper is to call out two facts about this approach
to one-dimensional (parallel) electron dynamics, and explore their
consequences. The first is that energy conservation and Liouville's
theorem guarantee that if the distant unperturbed electron
velocity distribution is stable and \emph{symmetric} (but not otherwise), then
the distribution within the wake is also stable. The second is that
parallel energy conservation is an \emph{approximation}, good only to
lowest order in the square-root of the electron to ion mass ratio
$\sqrt{m_e/m_i}$. Consequently the stability properties of the
electron distribution function are crucially determined by the mass
ratio, and finite mass ratio may need to be accounted for.

Moreover, theoretical models that use artificial values of mass
ratio closer to unity than in nature, which is common in PIC codes
applied to the moon wake \cite{Farrell1998,Birch2001,Farrell2008}, will violate parallel energy
conservation more strongly, and will lead to more unstable electron
distributions in wakes, thus failing to represent actual physics. 

It is shown that finite electron mass gives rise routinely to
bump-on-tail instability in quasi-neutral magnetized plasma wakes. The
quasi-linear strength of the instability is evaluated for different
ion to electron mass ratios. For physical values of the ratio (1835
for protons) the energy transferred from unstable electrons to plasma
waves is up to $10^{-3}$ of the total energy of the distribution
function. This level of instability is may well bear on space-craft
observations \cite{Kellogg1996,Halekas2011}. For artificial ion to
electron mass ratio of 25 (a value not infrequently used in PIC
simulations) instability energy fractions roughly 100 times higher
occur. These are unphysical.

\section{Self-similar potential solution}\label{simpotl}

We briefly summarize the standard self-similar solution arising from
the ion dynamics\cite{Gurevich1969,Sack1987,Manfredi1993}. It serves
to set the shape of the potential variation in which the electron
dynamics is analysed. Ignoring ion pressure (which is quite well
justified even when the external ion temperature is comparable to
$T_e$ because of the ions' cooling caused by their
acceleration\cite{Gurevich1969}) the ion continuity and (parallel)
momentum equations are sufficient. Using the scaled $x$-variable,
velocities normalized to the (undisturbed) sound speed, $c_s$, and
potential in units of $T_e/e$, the equations can be solved for
Boltzmann density variation to obtain, when $y/x \ge -1$:
\begin{equation}\label{ionsoln}
  \phi=-1-y/x\ ;\qquad n=n_\infty\exp\phi\ ;\qquad v_{\parallel i} =
  -\phi\ .
\end{equation}
For $y/x<-1$, the potential is undisturbed: $\phi=0$, $n=n_\infty$.  We
use subscript $\infty$ to denote the external, undisturbed values, and
we have set the external parallel drift to zero, $v_{\parallel
  i\infty}=0$. (See section \ref{appendix} for non-zero $v_{\parallel
  i\infty}$.)

\begin{figure}[htp]
  \hbox to \hsize{\hss
  \includegraphics[width=3.8in]{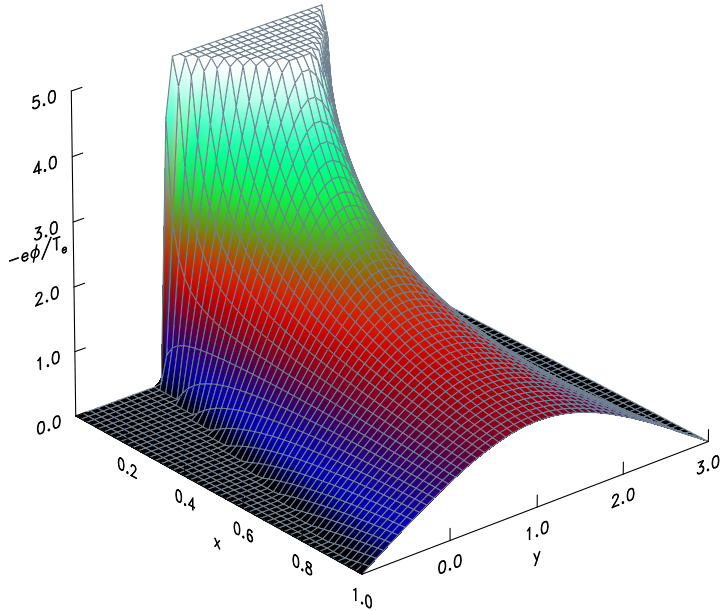}
  \hss}
  \caption{Approximate electron
    normalized potential energy ($-e\phi/T_e$) as a function of
    position, corresponding to eq.\ (\ref{stream2})\label{webpotl}}
\end{figure}
This self-similar solution, in which parameters are a function only of
the ratio $y/x$, holds only to the extent that ions arriving through
the wake from the other side of the object can be ignored. The form
applies with the substitution $y\rightarrow 2R-y$ (where $R$ is the
object's half-height) to the upper limb of the wake. See
Fig\ \ref{geom}. At the axis of symmetry, the two opposite ion streams
merge. Although the equations are then not rigorously justified, the
resultant can be reasonably approximated \cite{Gurevich1969} by taking
the density to be the sum of the stream given by eq.\ (\ref{ionsoln})
plus its equivalent with $y\rightarrow 2R-y$. The resulting potential
can quickly be shown to be
\begin{equation}\label{streams1}
\phi = -1 -R/x + \ln\left[2\cosh\left({R-y\over x}\right)\right] .
\end{equation}
However, this expression does not exactly go to zero at $y=-x$, so it is
better to subtract from it its value at $y=-x$ and use the resulting form:
\begin{equation}\label{stream2}
  \phi = \ln\left[\cosh\left({R-y\over x}\right)\right]
- \ln\left[\cosh\left(R+x\over x\right)\right] .
\end{equation}
The difference is negligible for $x\lesim 1$.  Fig.\ \ref{webpotl} is
a 3-D rendering of this potential with spatial units scaled so that
$R=1$.  [In the very far wake, at distances exceeding $Rv_\perp/c_s$
  ($x>1$), some simulations, e.g.\ \cite{Farrell1998}, indicate a more
  complicated wake potential structure. We are interested in the nearer wake
  where the electron instability processes are stronger and the
  potential structure more robust.]

For analytic convenience, two cruder approximations have also been
explored. The ``linear $\phi$'' is simply to suppose that
eq.\ (\ref{ionsoln}) applies up to the axis of symmetry and the upper
solution applies above it; the ``flat-top $\phi$'' is to adopt eq (2)
only until $-\phi$ reaches the value $1+R/x-\ln2$ and then flat
otherwise, in the near-axis region. For these cruder forms the orbit
can be integrated analytically which is useful for verification of the
numerical orbit solution to be described in section \ref{nonconserv}.

\section{Electron stability with energy conservation}\label{drift}

The instability we consider here is that of electrostatic waves
arising from the parallel electron distribution function shape. Other
possible mechanisms include anisotropy-driven instabilities involving the
magnetic field and instability arising from the
two-stream nature of the ions, which has been characterized
elsewhere\cite{Gurevich1969,Gurevich1975}. We ignore these other instabilities and focus on the
electrostatic electron instability which will be by far the fastest
growing, if it exists. The Penrose criterion states that instability arises if
and only if there is a minimum in the one-dimensional distribution
function $f(v)$ at a velocity $v_0$ and that
$\int[f(v)-f(v_0)]/(v-v_0)^2 dv > 0$ \cite{schmidt79}. It therefore suffices for
stability to demonstrate that $f(v)$ is monotonically decreasing
either side of a single maximum. Maxwellian or Kappa distributions,
even with a shift of velocity origin representing parallel drift, are
stable by this criterion.

At any point in space, the collisionless electron velocity
distribution function at any velocity $v_\parallel$ may in principle
be found by tracking backwards along the (phase-space) orbit until one
arrives somewhere in the unperturbed plasma, where the distribution
function at the corresponding energy is known. If such an
orbit instead tracks back until it intercepts the object, then under
the assumption the object is purely absorbing, that orbit is
unpopulated.

Recall, however, that all the orbits move with a constant velocity in
the perpendicular direction (in the rest frame of the object). If
the typical electron thermal velocity is much larger than this
drift velocity, then the spatial trajectory of all the electron orbits
will be dominated by parallel motion rather than by the perpendicular
drift, and as a result hardly any orbits will actually
intercept the object.

In a wake, the electric potential is negative, repelling
electrons. The height of the potential energy hill that the electrons
encounter, which peaks at the axis of symmetry of the wake, depends
upon the  $x$-position in the wake. For a point on the
negative $y$ side of the wake, orbits with sufficiently negative
velocity at the point track backwards over the hill to the upper side
of the wake. Others are reflected by the hill (if $v_y<0$) and track
to the lower side, as illustrated in Fig.\ \ref{geom}. Nevertheless, if
the distant electron distribution is reflectionally symmetric about
the wake axis, and parallel energy is conserved, then it makes no
difference which side of the wake the orbit originated from. The
electron distribution at the point of interest is then equal to the
unperturbed distribution shifted by energy.

If, on the contrary, the distant electron distribution is asymmetric
in velocity, for example shifted in velocity, representing a net
parallel mean velocity of electrons, the electron distribution in the
vicinity of the wake will then possess a discontinuity at the marginal
velocity whose orbit \emph{only just} crosses the
potential hill. Higher energy electrons have orbits that are monotonic
in $y$ and have arisen from the negative $v_\parallel$ part of the
distribution in the upper region. Lower energies are reflected orbits
that arose from the positive $v_\parallel$ part of the distribution
function in the lower region. 
The distribution will have a
local minimum if the shift of velocity in the distant distribution is
in the negative direction. In other words, if the distribution of
electrons that cross the hill (i.e.\ have negative distant velocity) is
larger (at the same energy) than those that are reflected (i.e.\ have
positive distant velocity) instability may arise.  [Mathematically the
  discontinuity is an infinite gradient, but acts as if the sign of
  $f'$ has changed.]

\begin{figure}[htp]
\hbox{\hss
    \includegraphics[width=3.1in]{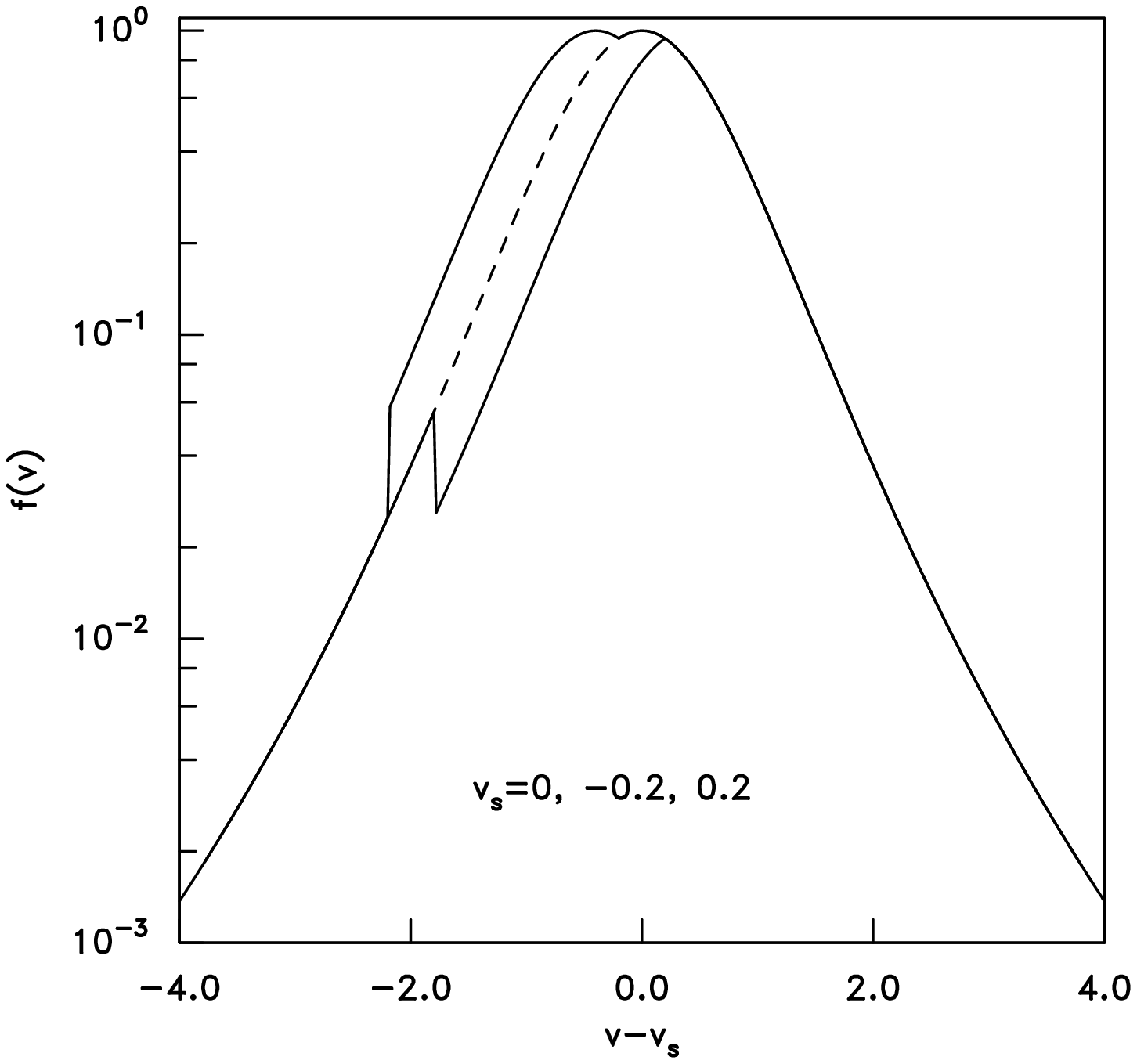}
\hskip-2.8in(a)\hskip2.65in
    \includegraphics[width=3.1in]{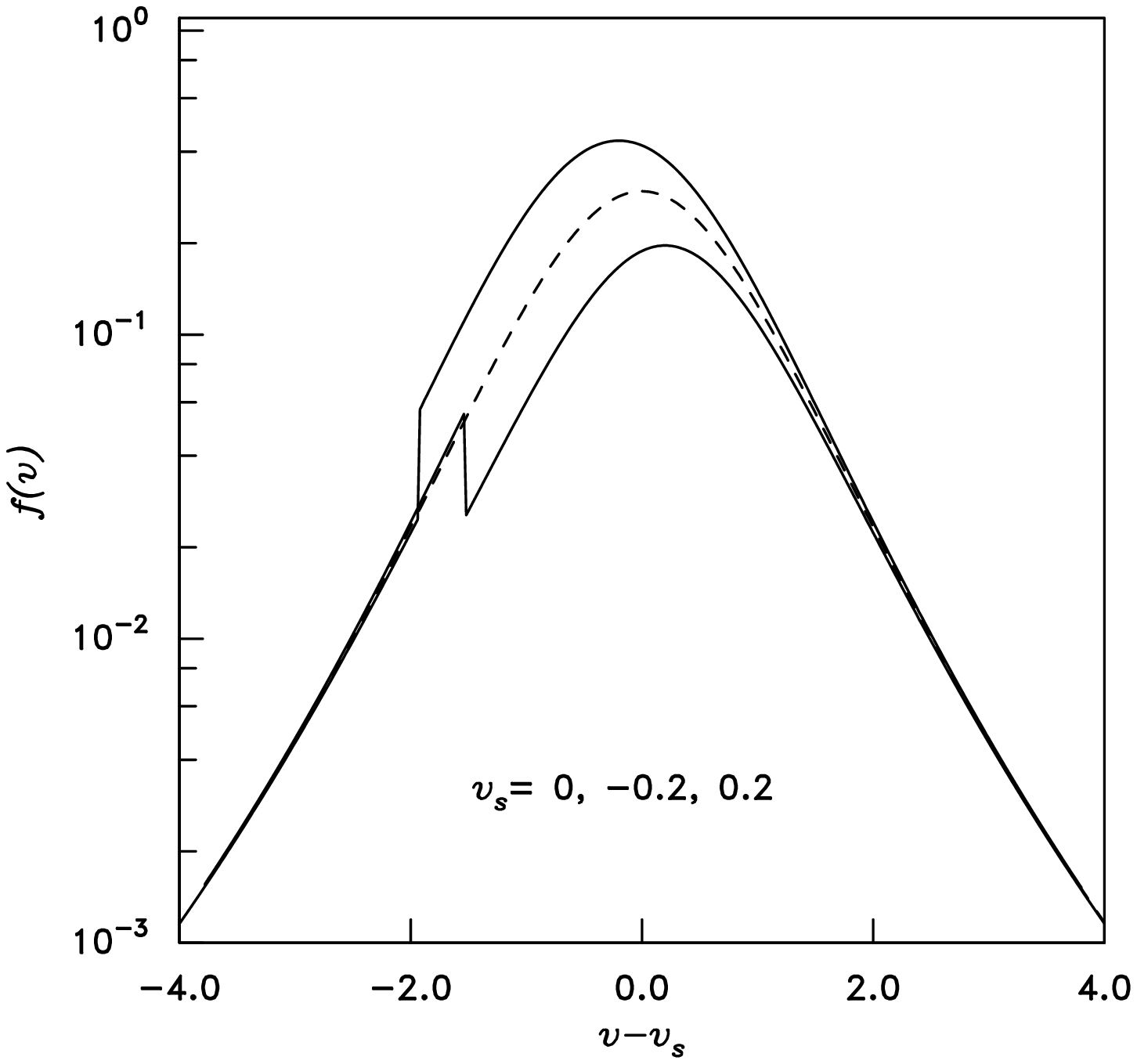}
\hskip-2.8in(b)\hskip2.65in
\hss}
  \caption{Distribution functions adjacent to a potential energy hill of
    normalized height $2$, when the external distribution has a velocity
    shift $v_S$. \label{shiftdist} (a) at potential zero, in the unperturbed
    region; (b) at a position where normalized potential is 0.5. The
    shape is unstable (bump-on-tail) if $v_s<0$.}
\end{figure}
Fig.\ \ref{shiftdist} illustrates the effect on the distribution
function, using a Kappa distribution to emphasize that the effect is
not dependent on Maxwellian distributions. The undisturbed electron
distribution is taken as $f(v)=
[1+(v-v_s)^2/(\kappa\theta^2)]^{-(\kappa+1)}$ with $\theta=1$,
$\kappa=2$, and the conserved energy is ${1\over2}v^2 + E_p$, where
$E_p$ is the potential energy in normalized units. The presence of the
potential hill forms an unstable distribution at the reflection
discontinuity if the distribution's velocity shift $v_s$ is negative,
so that reflected electrons (with $v$ immediately above the
discontinuity) have a smaller unperturbed distribution
function. Postive $v_s$ can give rise to a dimple at the top of the
distribution function (Fig.\ \ref{shiftdist}(a)), but this is unlikely
to be Penrose-unstable for small $v_s$, and disappears at positions
where $E> {1\over2}v_s^2$ (Fig.\ \ref{shiftdist}(b)).

Asymmetry in the electron distribution is known to arise in the solar
wind especially in the form of the ``Strahl'', an energetic parallel
electron tail flowing away from the sun, believed to arise because of
magnetic-mirror force acceleration. However, this population is
predominantly at high energies and has a relatively low density; so
the instability will saturate at a modest level.


\section{Instability from energy-nonconservation}\label{nonconserv}

We now consider an effect that can cause instability even
when the external electron distribution is reflectionally
symmetric. Therefore in this section only unshifted external
distribution functions are considered. The effect arises because
parallel electron energy is \emph{not} exactly conserved along
orbits. Energy non-conservation can be considered, in the moving frame
of the background drift, to arise from the fact that the wake
potential is changing with time. Alternatively, in the frame of the
object, it comes from the convective perpendicular velocity, and may
be understood from the parallel momentum equation in steady
state ($\partial/\partial t=0$)

\begin{equation}
  v_\perp {\partial v_\parallel \over \partial x} + v_\parallel
  {\partial v_\parallel \over \partial y} = {e\over m_e} {\partial \phi
      \over \partial y}
\qquad\longrightarrow\qquad
  {\partial v \over \partial x} + v
  {\partial v \over \partial y} = {1\over m_r} {\partial \phi
      \over \partial y}.
\end{equation}
The first form here is the dimensional equation to help the reader
with familiarity, the second is the equation expressed in
dimensionless units, where $m_r=m_e/m_i$, and the parallel subscript
has been dropped for brevity.  If the first term on the left hand side
of either of these equations were absent, then that side becomes
simply ${1\over2}\partial v^2/\partial y$, a total derivative, which
leads immediately to the conservation of energy ${1\over2} v^2-
\phi/m_r=constant$, in normalized units. The extra term means parallel
energy is not conserved. The extra term is apparently smaller than the
other terms by a factor of order $v_\perp/ v_e$, although we shall see
in a moment that the factor is actually $c_s/v_e= \sqrt{m_r}$, but it
is not immediately obvious how important it is,
and what it actually does to the distribution function.

\begin{figure}[htp]
\hbox to \hsize{\hss
    \includegraphics[width=3.6in]{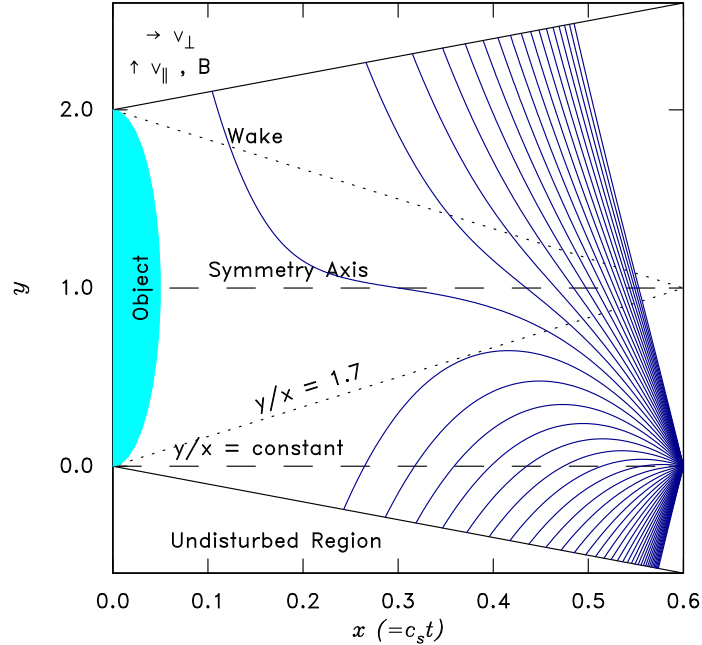}
\hss}
  \caption{Orbits that all end at the point (0.6,0.)\ determining
    the distribution function there. Mass ratio
    $m_r^{-1}=25$.\label{fancyorbits}}
\end{figure}

Heuristically one can understand how a depression in the electron
distribution function (and hence an instability) arises as
follows. Figure \ref{fancyorbits} plots a series of actual orbits
for illustration.
The electron orbits that are most affected by the convective drift are
those which spend longest near the peak of the potential (i.e.\ the
symmetry axis). They do so because their parallel velocity is near
zero there. These are the \emph{marginal} orbits that just barely make
it over the potential hill or are just barely reflected. They start in
the unperturbed region at a parallel speed that has to be high enough
to climb the potential hill at a position where the hill is high
(because $x$ is smaller). They drift across the field during
their time near the potential peak (not gaining parallel energy) to
where the potential is lower (because $x$ is larger). Then their
evolving parallel speed carries them down the hill to the final
position, but as it does so they gain parallel energy corresponding
only to the difference between the lower potential peak and the final
position. The distribution function at the final position and speed
($f(v)$) is
equal to the external distribution function at the starting point with
the starting speed ($f_\infty(v_\infty)$), but the starting speed is higher than
would be the case with energy conservation. This effect is present for
all orbits but it is much stronger for marginal orbits. The
distribution function $f$ is therefore smaller for marginal orbits
(because their starting speed is higher) than it is for orbits further
from marginal. That is, a depression is formed near the marginal
velocity.

\subsection{Analytic Orbit Solution}
To quantify the effects of the energy-nonconservation, one must solve
the orbit equation. This can be done analytically when the potential
has the form that arises from the self-similar solution of the ion
problem (eq \ref{ionsoln}). One should recognize that to adopt the ion
solution form of potential is only an approximation. We are
calculating distributions that are not exactly those giving the
Boltzmann or polytrope relationship between $\phi$ and density,
assuming the effect on that relationship is small. An iterative
approach to the solution could of course in principle solve for the
self-consistent potential incorporating the full numerical electron
distribution (and also the effect of the overlap of the ion
streams). But that would be a far greater task, and yield little extra
insight for the electron distribution stability. We will therefore be
content with observing after the fact that the electron density
deviates only negligibly from its assumed relationship, at least in
those regions where we have not made other approximations of
comparable significance.

We work henceforth in dimensionless terms. Writing $z\equiv
y/x$, suppose $\phi = \phi(z)$ is a quadratic in $z$, so that
\begin{equation}\label{phigrad}
  {d\phi\over dz} = A + Bz ,
\end{equation}
where $A$ and $B$ are constants. For the Boltzmann-density case,
actually $A=-1$, $B=0$. For for the polytropic problem, arising from
Kappa-distribution electrons, form (\ref{phigrad}) is also obtained, but with
$B\ne 0$. Also, the orbit is
\begin{equation}
  {dy\over dx} = v \qquad \left(\Rightarrow\quad 
{d\over dx}={\partial\over \partial x} + v{\partial \over \partial y}
\quad\mbox{and}\quad
{dz\over dx} = {v-z\over x}\ \right)
\end{equation}
and of course
\begin{equation}\label{vxorbit}
  m_r {dv\over dx} = {d\phi\over dy}.
\end{equation}
So we have
\begin{equation}
x {dv\over dx}=
 (v-z) {dv\over dz} = {x\over m_r } {\partial\phi\over \partial y} = {1\over m_r} {d\phi\over dz} = (A + Bz)/m_r .
\end{equation}
When B is non-zero, the equation $(v-z)dv/dz=(A+Bz)/m_r$ can be integrated by rendering it
into homogenous form and separating the variables through the substitutions
$s=z+A/B$, and $u=(v+A/B)/s$. The final result is
\begin{equation}\label{noncons}
  [v+A/B-(z+A/B)u^+]^{P^+}[v+A/B-(z+A/B)u^-]^{P^-} = const
\end{equation}
where
\begin{equation}
  u^\pm = (1\pm\sqrt{1+4B/m_r})/2\ ,\qquad \mbox{and} \qquad 
P^\pm = (1- {\pm1\over \sqrt{1+4B/m_r}})/2 \ .
\end{equation}
Eq.\ (\ref{noncons}) is the replacement for the conservation of
energy, in our convecting situation.
For the Maxwellian case, $B=0$, $A=-1$, the integration proceeds more easily
to obtain
\begin{equation}\label{nonconm}
  m_r v - \ln( m_r[v-z] +1) =const.
\end{equation}
 One can verify for either of these conservation equations
 (\ref{noncons},\ref{nonconm}) that to lowest order in $\sqrt{m_r}$ as
 $m_r\to 0$ they become ${1\over 2}m_rv^2 - (Az+Bz^2/2)= const$, which
 is exactly energy conservation.  For any orbit that moves only in the
 region where the one-sided self similar potential eq. (\ref{ionsoln})
 applies, because it is directed inward towards the potential energy hill or was
 reflected well away from its peak, these equations apply. In that
 case, the orbit at phase space position $(z,v)$ tracks back to some
 corresponding point in the lower undisturbed plasma
 $(z_\infty,v_\infty)$ and $f(z,v)=f(z_\infty,v_\infty)$. (We discount
 for now orbits that might reach the object, which are unpopulated.)
 Then, provided that the function $v_\infty(v)$ is monotonic, if
 $f(z_\infty,v_\infty)$ has no minimum then neither does
 $f(z,v)$. This part of the electron distribution is stable. [This
   point contradicts the heuristic arguments of \cite{Farrell2008}
   which claimed that ``time-of-flight processes'' form ``an inward
   directed ... electron beam''. No inward ``beam'' in the sense of a
   secondary maximum of the electron distribution function can form by
   such processes.]

However, for orbits that move close to the axis of symmetry, or over
the potential energy hill from the top to the bottom, more elaborate
analysis is required, because eq.\ (\ref{ionsoln}) does not apply. It
proves possible to extend the analytic treatment for the two cases
described in section 2, by joining solutions in different regions:
above and below the symmetry axis, or at the edge of the flat
potential region.

To accomplish the joining, one must integrate the orbit in $x$ and $y$
(not just the self-similar coordinate $z$). The orbit can be expressed
as a single (complicated) quadrature for the polytrope case. But since
the form of the distribution does not qualitatively alter the effect,
only the analytically simpler solution for a Maxwellian distribution
is given here.  In that case the orbit eq.\ (\ref{vxorbit}) is simply
$m_rdv/dx=-1/x$, with the immediate integral $m_rv+\ln x =
const$. Integrating again and requiring the velocity to be $v_0$ at
the final point $(x_0,y_0)$ we find the solution
\begin{equation}\label{morbit}
  y=y_0+x\ln(x_0/x)/m_r + (v_0+1/m_r)(x-x_0).
\end{equation}

For the ``linear'' and ``flat-top'' potential forms, the full orbit
can then be constructed by joining solutions like this
(eq.\ \ref{morbit}) with straight-line sections in which the potential
gradient is zero. They give rise to unstable electron distributions.

\subsection{Numerical Orbit Solution}

For the potential form of eq (\ref{stream2}) it is not
possible to solve for the orbit analytically. So instead, a computer program has
been implemented to solve for the orbit $y(x)$ by integrating backward
from the end-point using a fourth order
Runge-Kutta integrator. This integrator has been benchmarked against
the analytic solutions (using the appropriate potential forms) giving
negligible systematic error, though noise arising from rounding errors
is slightly worse. All subsequent results shown use  eq.\ 
(\ref{stream2}) for the potential, as the appropriate approximation
for interpenetrating ion streams. 

Tracking back many such orbits for different end-point velocities
provides the electron velocity distribution function at the end-point
of interest $(x_0,y_0)$. The parameters that govern the result are
just that point position and $m_r$.  It is convenient to measure $y$
in units normalized to the object half-height, $R$. Then the units of
$x$ are $R v_\perp/c_s$: i.e.\ larger by the drift Mach number.

Fig.\ \ref{fancyorbits} illustrates with a restricted number of
velocities a case with end-point closer to the object ($x_0=0.6$) and deeper
into the wake ($y_0=0$) than Fig.\ \ref{geom}. And
Fig.\ \ref{fancyorbits} has a mass ratio $m_r^{-1}=25$, characteristic
of a non-physical calculation with enhanced electron mass. The
electron orbits extend backwards most of the way to the object. The
orbits bifurcate when they make the transition from crossing the
symmetry axis to being reflected before it. Since the bifurcation
occurs at the marginal orbit, we can graphically summarize the orbit
behavior for the far larger number (typically 500-1000) of orbits
actually used for distribution function evaluation simply by plotting
two marginal orbits.
 \begin{figure}[h,t,p]
     \hbox to \hsize{\hss
     \includegraphics[width=3.1in]{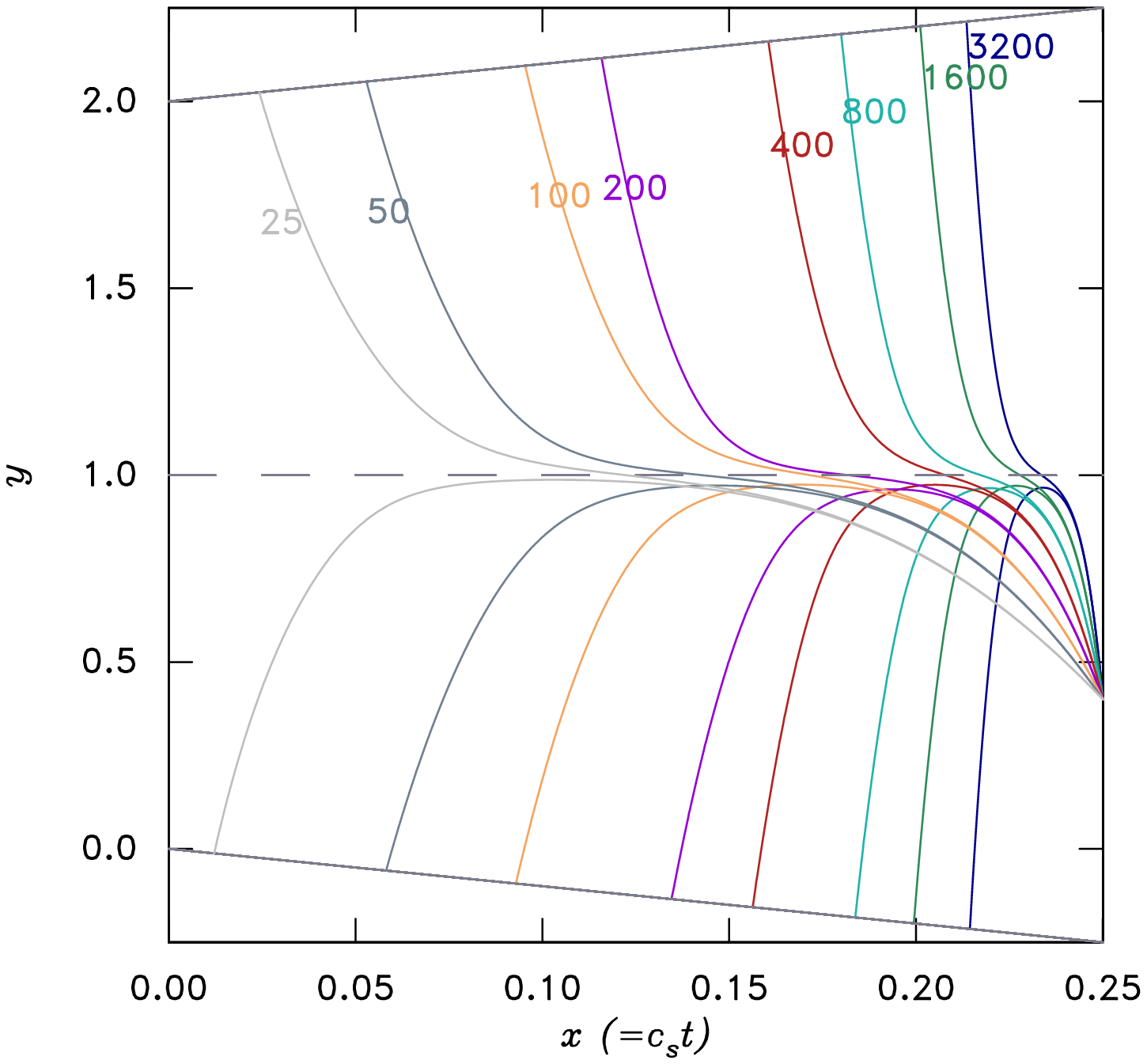}
     \hskip-2.8in(a)\hskip2.65in
     \includegraphics[width=3.1in]{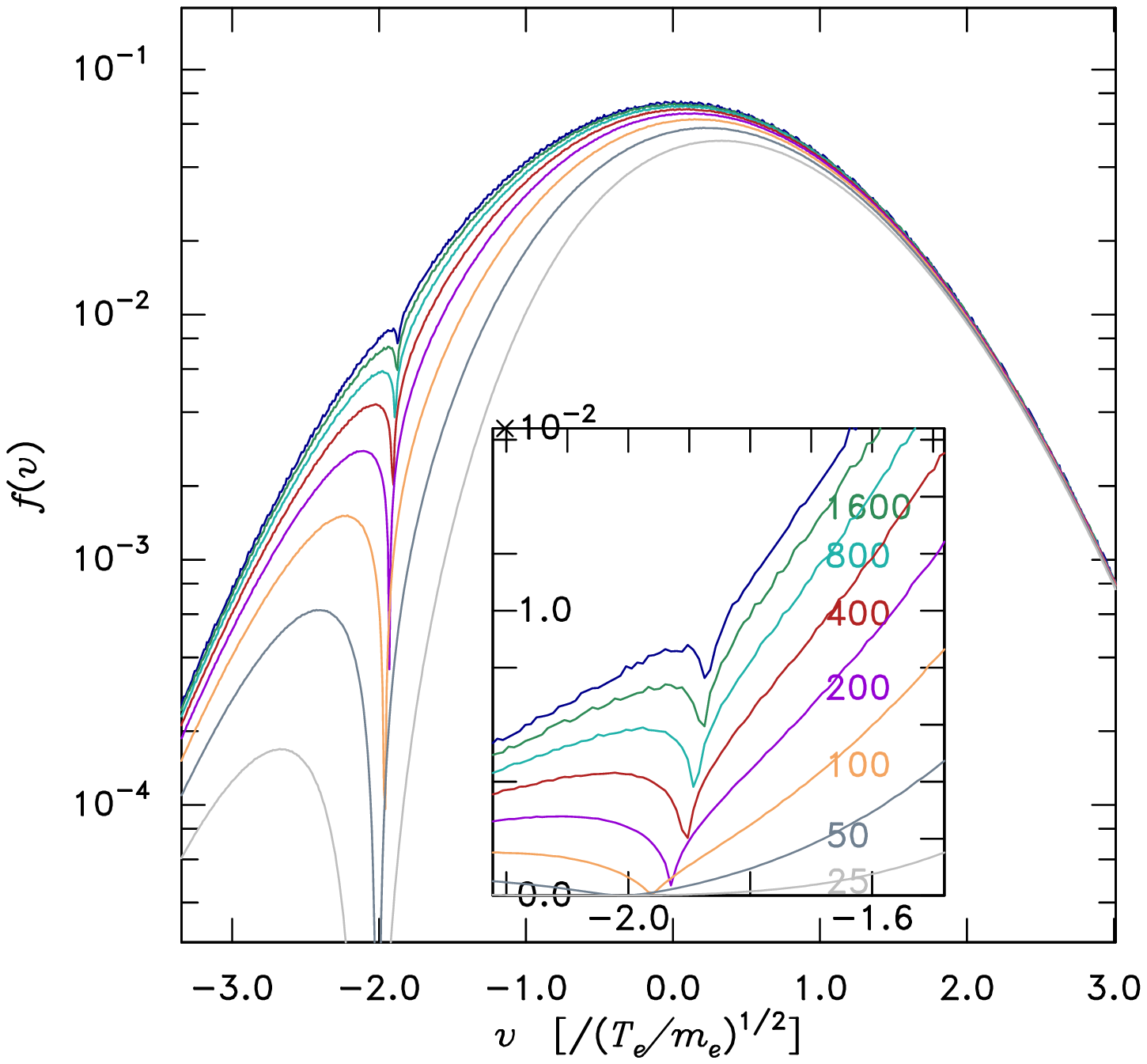}
     \hskip-2.8in(b)\hskip2.65in\hss}
   \caption{(a) Marginal orbits for a range of mass ratio
     $m_i/m_e=m_r^{-1}=$ 25, 50, 100, 200, 400, 800, 1600, 3200.  (b)
     Electron distribution functions at the final point (0.25,0.4),
     for these parameters. The insert expands the display of the
     unstable region of velocity space, plotting $f(v)$ on a linear
     scale.\label{orbits26} }
 \end{figure}

In Fig.\ \ref{orbits26} are shown marginal orbit examples for a range
of $m_r^{-1}$ values. The reduction of mass ratio leads to increasing drift
effect. Tracking back the marginal orbits, their starts are
increasingly closer to the object. The resulting distribution
functions are plotted in Fig.\ \ref{orbits26}(b). Lower mass-ratio cases show
wider and deeper distribution minima. Formally, all the
distributions of Fig.\ \ref{orbits26} are Penrose unstable. The linear
inset close-up of the marginal velocity region clearly shows a minimum
in $f(v)$. Although it arises from effects that cause a difference
between reflected and unreflected orbits its occurrence requires no
asymmetry in the distant distribution. The bump-on-tail of the case
corresponding to nature ($m_i/m_e\sim 1600$) is fairly small and the
minimum narrow, compared with the enhanced-electron-mass cases.

\begin{figure}[htp]
    \hbox to \hsize{\hss
    \includegraphics[width=3.1in]{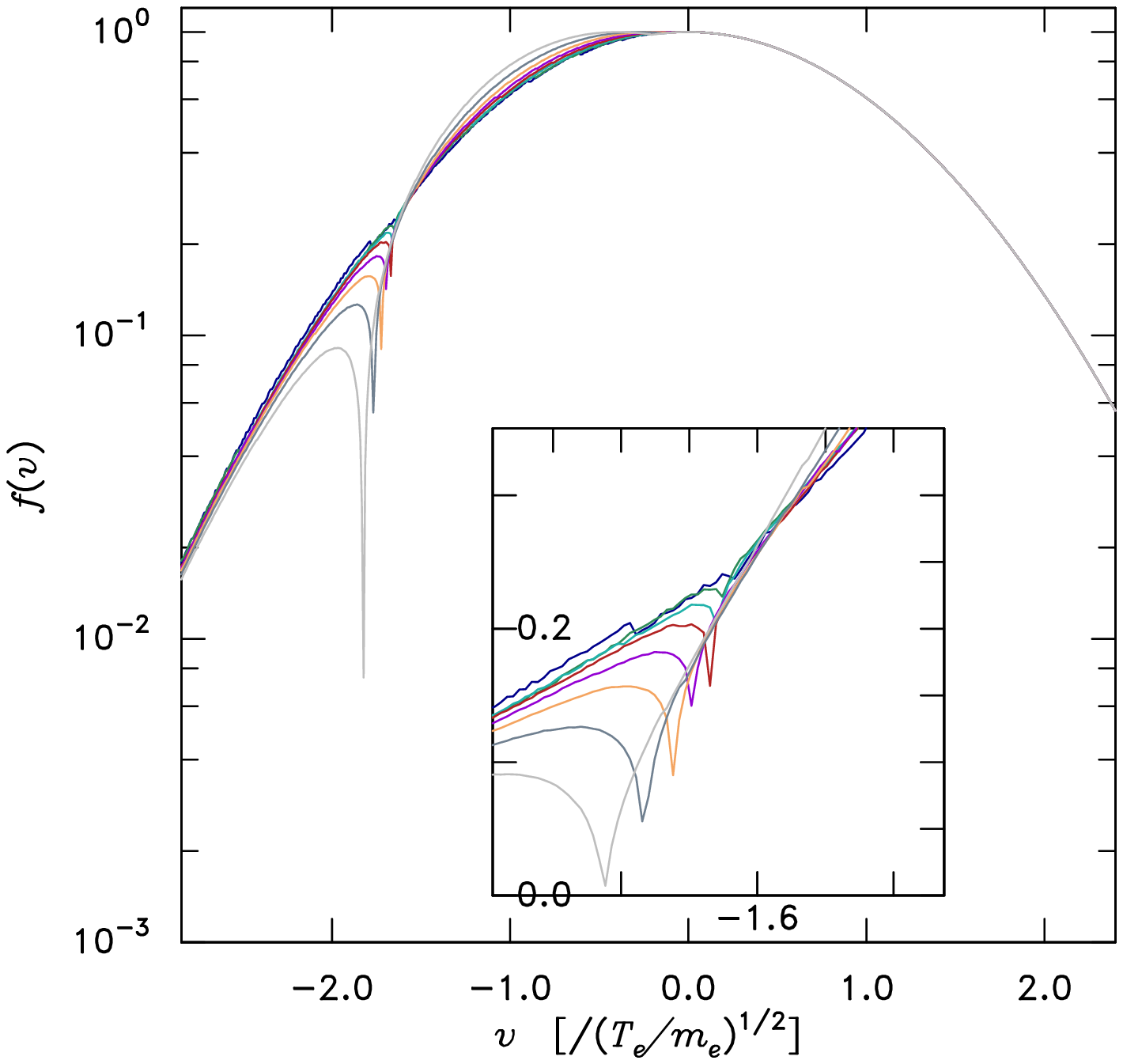}
    \hskip-2.8in(a)\hskip2.65in
    \includegraphics[width=3.1in]{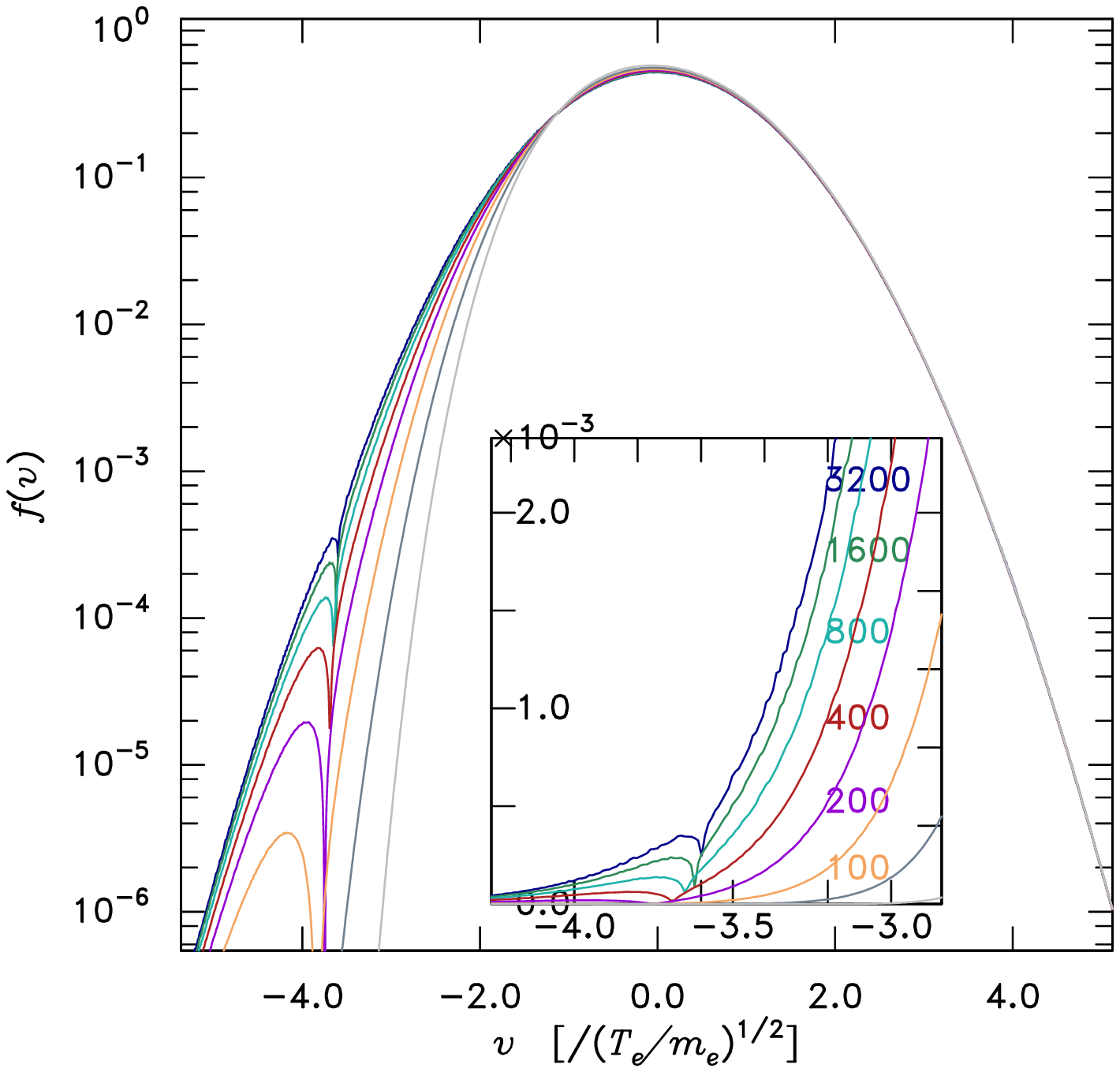}
    \hskip-2.8in(b)\hskip2.65in
    \hss}
  \caption{Velocity distributions for a range of mass ratio
    $m_i/m_e=m_r^{-1}=$ 25, 50, 100, 200, 400, 800, 1600, 3200 . (a)
    Final point (1.0,-1.05). (b) Final point (0.15,-0.05).\label{orbit15} }  
\end{figure}
If we choose a point out on the edge of the wake region or even
outside it, such as is shown in Fig. \ref{orbit15}(a)
for position (1.0,-1.05), then there is no suppression of the peak of
the distribution. It is equal to unity, the normalized value in the
external plasma. At this position, while a deep hole is present in $f(v)$
for large electron mass, the distribution at physical mass ratio is
practically stable within the noise level of the calculation.

In contrast, when the point of interest moves closer to the object
(smaller $x$) as
in Fig.\ \ref{orbit15}(b), the marginal velocity is further out on the
tail of distribution function. An unstable minimum is present; but it
is at a distribution height that is very small, $\sim 10^{-4}$ of the
peak. As $x$ is decreased still further the instability strength
is eventually actually reduced to a negligible level as the gaussian
tail decay becomes predominant.

These velocity trends arise, of course, because the height of the
potential hill at position $(x,1)$ is equal to $\ln(\cosh(1+1/x))$, which
approximately determines the corresponding marginal ($f$-minimum) velocity
$v \approx 2\sqrt{(T_e/m_e)\ln(\cosh[(1-y)/x])}$.

\section{Nonlinear perturbation magnitude}
To quantify the instability's significance for unstable electron
distributions it is helpful to use a quasilinear estimate of the
final state of the distribution function after the instability has
grown and saturated. This is based upon the approximation illustrated
in Fig.\ \ref{quasilin}.
\begin{figure}[tp]
    \hbox to \hsize{\hss
      \vbox{\hsize=2.3in\includegraphics[width=2in]{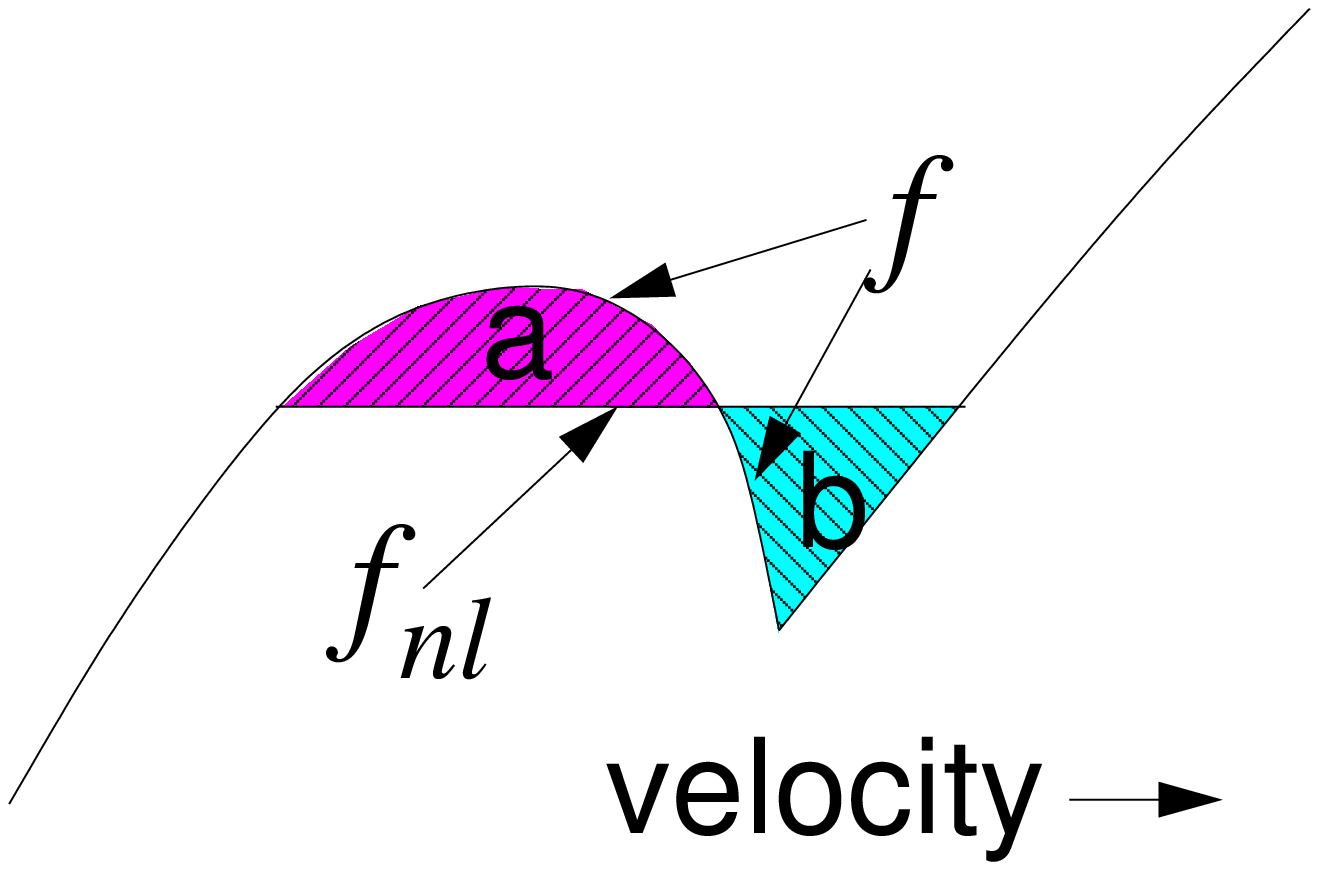}\leavevmode\vskip
        1cm}
    \hskip-2.3in(a)\hskip2.65in
    \includegraphics[width=3.3in]{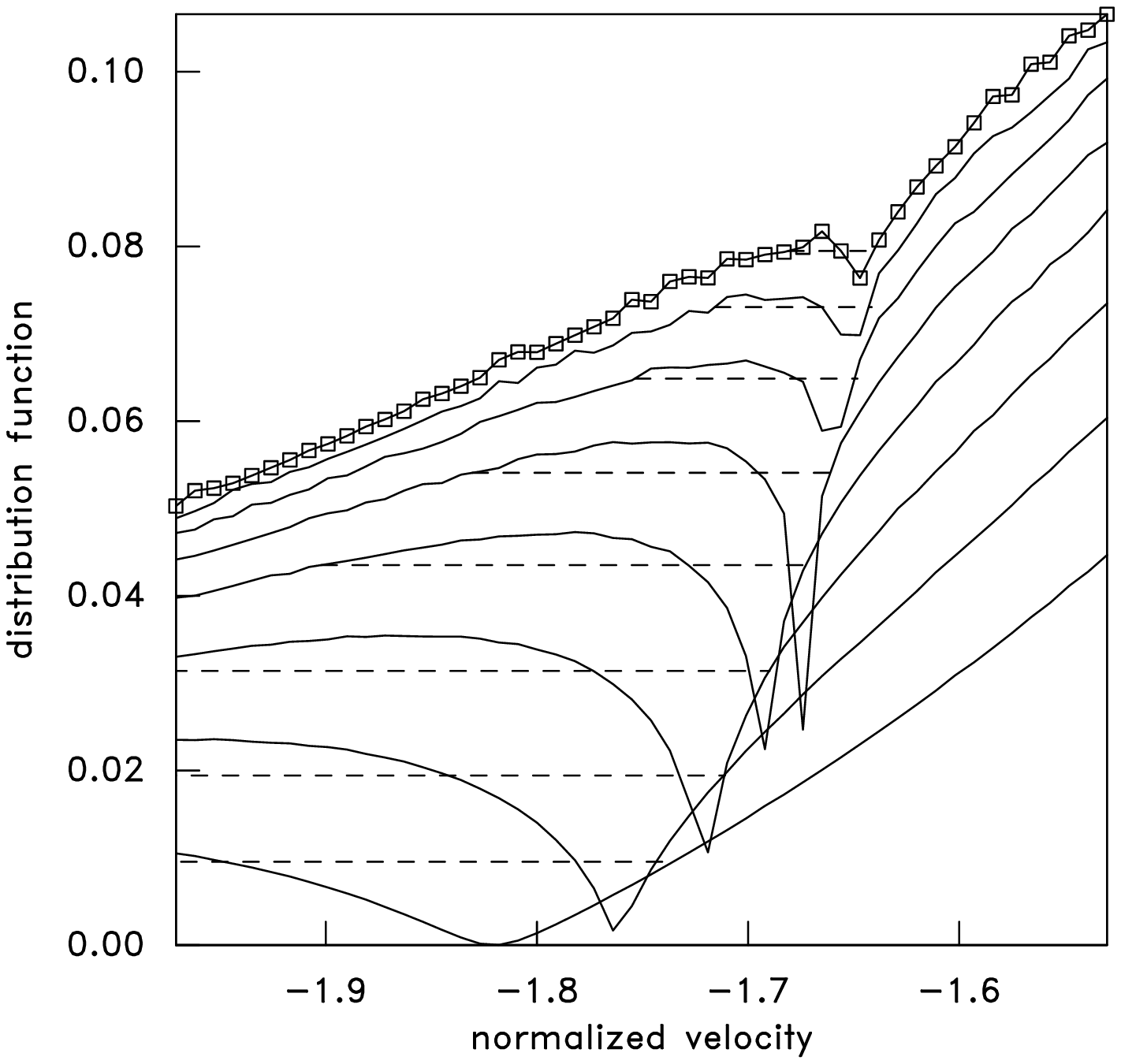}
    \hskip-2.8in(b)\hskip2.65in
    \hss}
  \caption{(a) Quasi-linear evolution of the distribution function is
    assumed to be from the initial state, $f$, to a final
    distribution, $f_{nl}$, where the unstable region is flattened by
    mixing. That flat region is taken to connect continuously to the
    initial distribution at its ends, and to conserve
    particles. (Area a = Area b.) (b) A plot of the actual numerical flattening
    process for distributions with     $m_i/m_e=m_r^{-1}=$ 25, 50,
    100, 200, 400, 800, 1600, 3200, and final position (0.6,0). The
    flattened plateau is shown by the dashed line. Individual velocity
    points are shown for the uppermost case
    (3200) to indicate resolution.  \label{quasilin}}
\end{figure}
The distribution is presumed to flatten in its unstable region by
quasilinear diffusion, conserving particles.  It is taken to connect
continuously to the unperturbed distribution at the edge of the region
of flattening. This specification uniquely defines the final state,
whose energy is lower than the initial state. The energy loss from the
resonant particles ($\Delta{\cal E}$) can readily be evaluated by
integration. It generally goes equally into wave energy and heating of
the bulk electron distribution\cite{davidson72}. So it represents
approximately twice the saturated turbulent wave energy expected to be
induced by the instability. The ratio of the resonant particle energy
loss to the total thermal energy (${\cal E}_0$) of the pre-flattening
electron distribution, gives a useful quantitative measure of the
strength of the instability.  The rounding error of the present
calculations becomes increasingly dominant below $\Delta{\cal E}/{\cal
  E}_0 \lesim 10^{-6}$, so distributions giving less than that cannot
be accurately assessed here. The background Langmuir turbulence
present in the solar wind is observed to be up to $\sim 10^{-6}$ of
${\cal E}_0$ \cite{Alpert1983}, so where numerically significant
resonant energy loss is found, it is well above levels in the
unperturbed solar wind.

For each distribution function, of the type illustrated in
Figs.\ \ref{orbits26}, \ref{orbit15} (but for a single specific mass
ratio, $m_r$) the values of $\Delta{\cal E}$ and ${\cal E}_0$ are
calculated using direct numerical integration. We perform a large
number of such calculations over a 20 by 20 grid of positions
throughout the $x$$y$-plane and display the results as a contour plot
of $\log_{10}(\Delta{\cal E}/{\cal E}_0)$ in Fig.\ \ref{cont25}.

\begin{figure}[htp]
    \hbox to\hsize{\hss
    \includegraphics[height=2.75in]{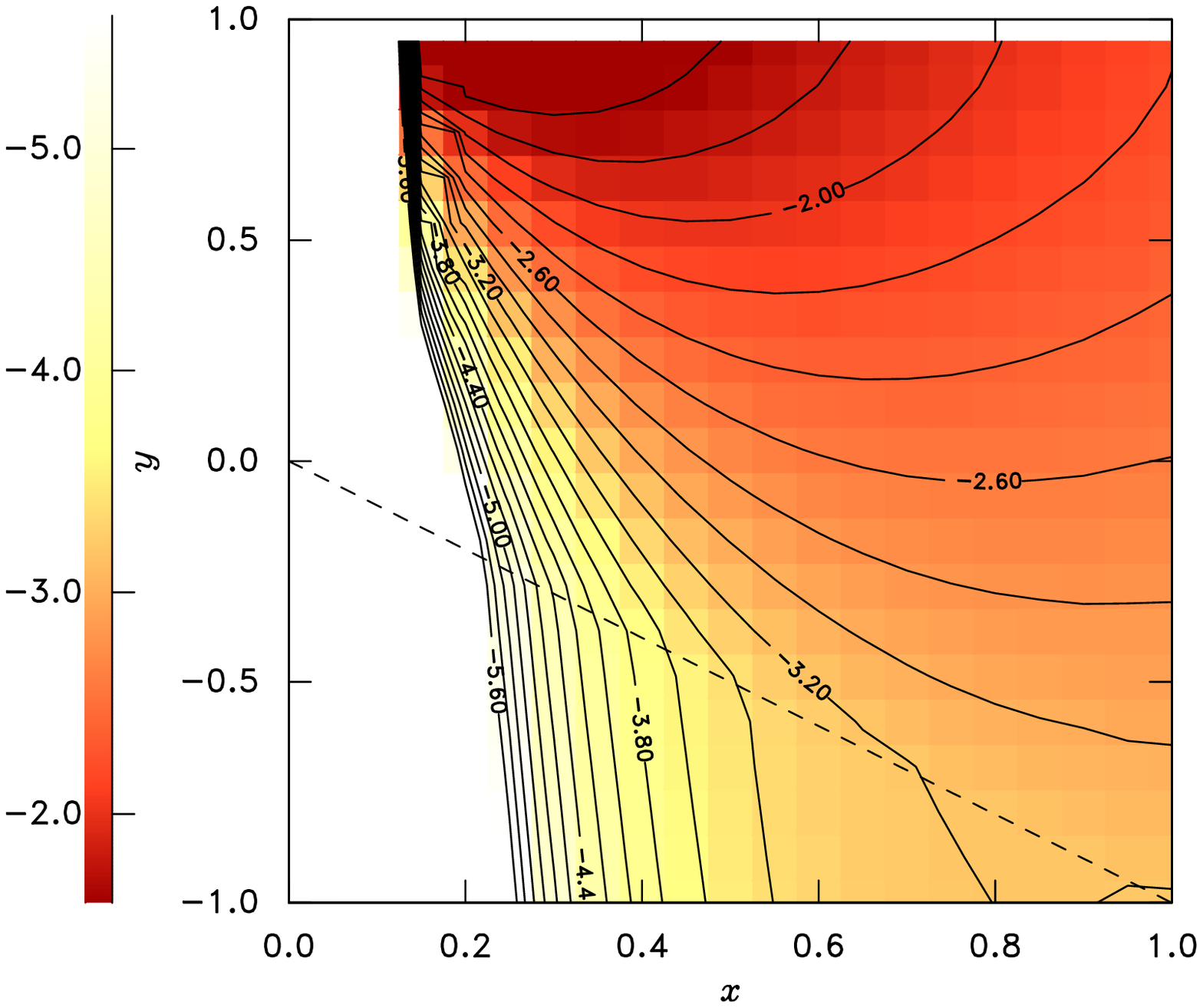}
    \hskip-2.8in(a)\hskip2.65in
    \includegraphics[height=2.75in]{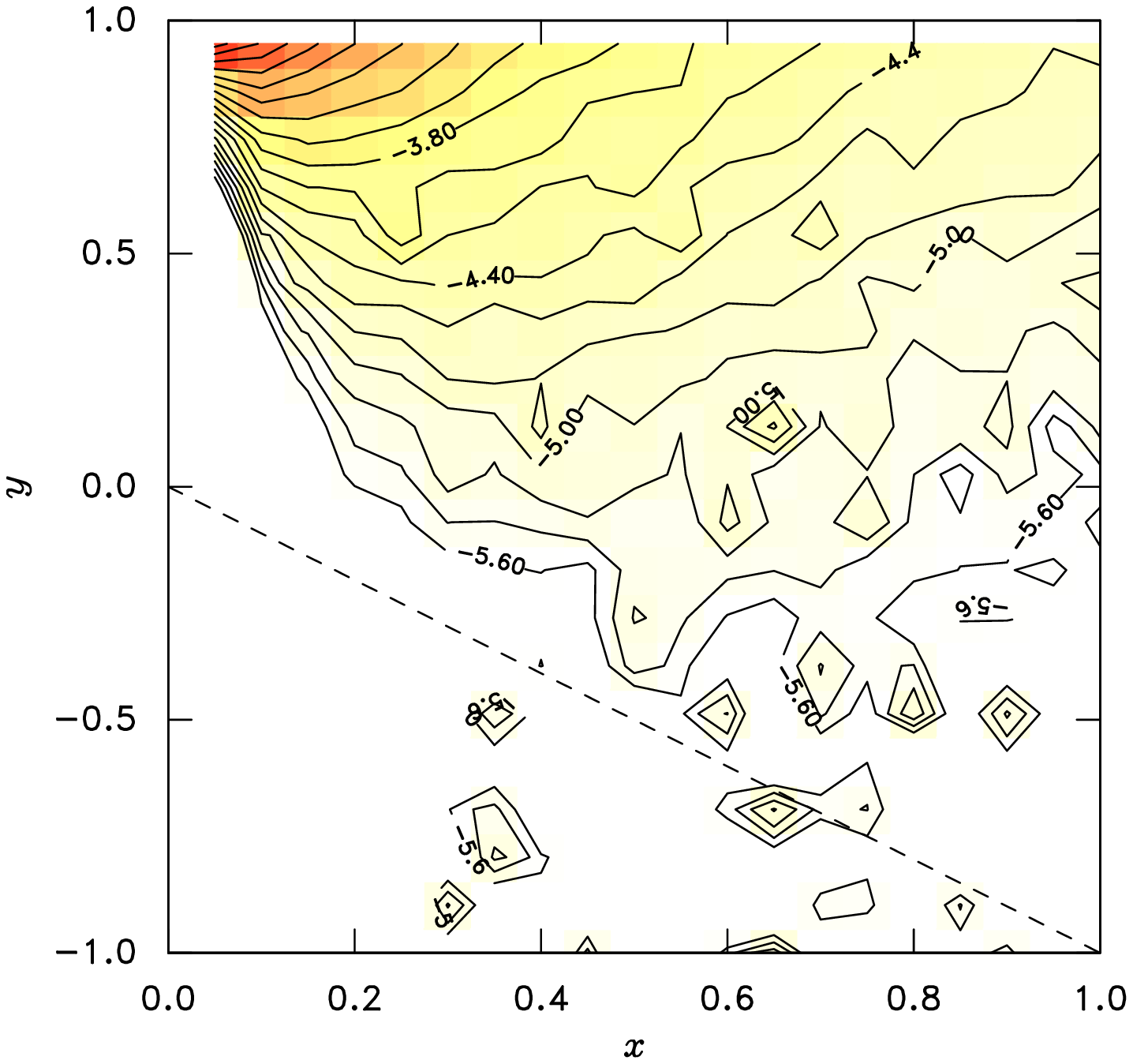}
    \hskip-2.8in(b)\hskip2.65in\hss}
  \caption{Contours (spaced by 0.2) of the relative instability
    strength $\log_{10}(\Delta{\cal E}/{\cal E}_0)$ over the wake
    $xy$-plane. (a) $m_i/m_e=25$, (b) $m_i/m_e=1835$. Geometry as in
    Fig.\ \ref{geom}. The $y$-units are object-radii, and the
    perpendicular distance in object-radii is equal to the
    perpendicular Mach number ($v_\perp/c_s$) times
    $x$.\label{cont25}}
\end{figure}

We observe that with artificially enhanced electron mass, so that
$m_i/m_e=25$, most of the wake region is unstable
(Fig.\ \ref{cont25}(a)). Indeed, that instability extends into the
external region where potential is undisturbed (i.e.\ to $y<-x$). The
only positions that are not unstable are at very small $x$. There the
distribution function is completely depleted at the marginal velocity,
in the way illustrated by Fig.\ \ref{orbit15}(b). The strongest
instability occurs near the wake symmetry axis ($y=1$) where a
substantial bump-on-tail occurs like that illustrated in
Fig.\ \ref{orbits26}. The total electron density (and energy density) itself
is also substantially depleted there, which contributes to the
enhancement of the \emph{relative} resonant energy loss by lowering
${\cal E}_0$.

In contrast, for realistic mass ratio $m_i/m_e=1835$
(Fig.\ \ref{cont25}(b)) the instability strength is greatly reduced:
by roughly two orders of magnitude. (Fig.\ \ref{cont25} is of the
\emph{logarithm} of instability strength.) The qualitative spatial
distribution of instability strength is fairly similar to
Fig.\ \ref{cont25}(a) but because it is quantitatively so much
smaller, it reaches approximately the noise level at the edge of the
wake's perturbed potential region. Roughly speaking, the instability
is significant for $y>0$, i.e.\ throughout the geometric wake.

Intermediate mass ratios give contour plots intermediate between the
two shown. The trend is illustrated in Fig.\ \ref{mrat} for the
instability strengths at several fixed points.
\begin{figure}[htp]
  \hbox to \hsize{\hss
  \includegraphics[width=4in]{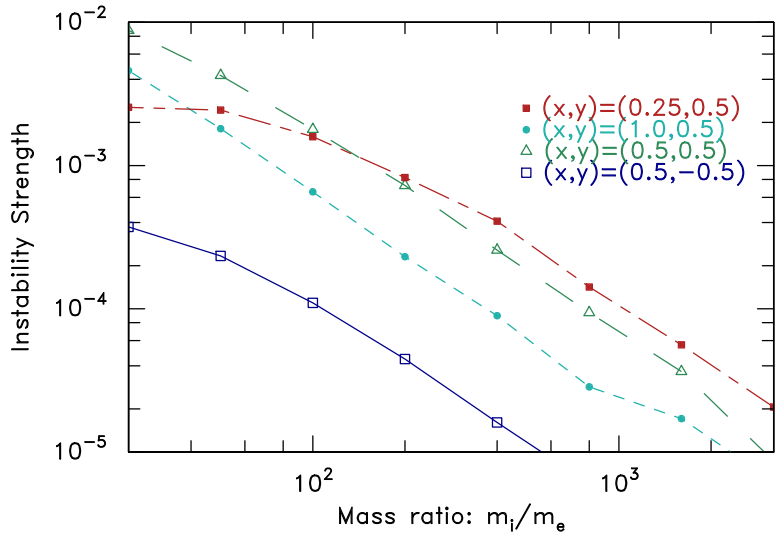}
  \hss}
  \caption{Variation of instability strength, $\Delta{\cal E}/{\cal
      E}_0$, with mass ratio for several end-points.\label{mrat}}
\end{figure}
The curves of $\Delta{\cal E}/{\cal E}_0$ versus $m_i/m_e$ fall almost
linearly in this logarithmic plot as $m_i/m_e$ increases, with slope
slightly steeper than -1.  There is no reduced threshold at which a
mass ratio is sufficiently large to give quantitative results
comparable to nature. One simply has to use the correct mass ratio.

The analytic potential approximations of ``linear $\phi$'' and
``flat-top $\phi$'' have also been explored to determine their
instability strength. It is found that the linear $\phi$ gives
substantially weaker instability, and the flat-top gives substantially
stronger instability (by factors between 10 and 100). This observation
demonstrates that the spatial profile shape of the potential plays a
major role in determining the strength of the instability. Flat-top is
more unstable because the marginal orbits spend more time near the
potential peak. Linear is less unstable because they spend less.  This
effect is sufficiently strong that the uncertainties in the potential
profile shape arising from the various approximations of the ion
problem solution may significantly affect the numerical values of the
instability strength. Therefore while the contour plots shown give a
correct order of magnitude instability strength, they cannot be
considered precise.

One should note that, because it is expressed in scaled units,
Fig.\ \ref{cont25} is essentially a \emph{universal} figure. The value
of the instability strength is not dependent upon plasma parameters
such as density or temperature so long as the Debye length and Larmor
radius are small compared with the object. Nor does it depend on
object size or drift Mach number. It does require the external electron
distribution to be well represented by a Maxwellian. Naturally
non-thermal external electron velocity distributions such as might be
represented by kappa-distributions, \emph{will} affect the instability
strength. However, the mechanism by which instability arises is the
same no matter what the external distribution is; and, notably, it
does not require non-thermal electron distributions.

For comparison, the values of the fractional quasi-linear energy loss
$\Delta{\cal E}/{\cal E}_0$ for the unstable distributions arising in
the energy-conserving case with external distribution shift,
Fig.\ \ref{shiftdist}, are (a) $8.5\times10^{-4}$, and (b)
$1.8\times10^{-3}$. Thus, inside the wake, the instability arising
from energy non-conservation is of the same order of magnitude as
would arise from a major external electron velocity shift: 0.2 times
thermal. The level of electric field fluctuation energy in thermal
equilibrium is ${\cal E}_0$ multiplied by a factor $\sim
1/(n_e\lambda_{De}^3)$. That factor, for the solar wind, is of order
$10^{-10}$. So all the turbulent energy levels discussed here are many
orders of magnitude higher than thermal.

\section{Oblique magnetic-field/drift}\label{appendix}

When the magnetic field and the drift velocity (of the plasma past the
object) are not perpendicular, the solutions we have presented still
apply immediately when interpreted in a way that this section
describes. We continue to use coordinates in which the magnetic field
is in the $y$-direction.  The drift velocity is now oblique in the
$xy$-plane. (This is a different choice of coordinates from what is
generally adopted in the space-physics community when discussing
wakes; they take the $x$-axis along the drift velocity). In these
magnetic-field coordinates, an oblique external drift is completely
equivalent to prescribing that in addition to the fixed perpendicular
drift velocity $v_\perp$ (in the $x$-direction), the ions in the
external region have a non-zero parallel velocity relative to the
object $v_{\parallel\infty}$
\cite{Hutchinson2008a,Hutchinson2008b}. Since the quasi-neutral
equations are entirely hyperbolic, as we have noted earlier, the
$x$-coordinate is equivalent to the time, $t$. In effect, the abscissa
of our plots can be considered either the distance from the object in
its frame of reference, or the time since passing the object in the
frame of the perpendicularly drifting plasma. In a frame of reference
moving at speed $v_{\parallel\infty}$ in the $y$-direction with
respect to the object, the external parallel ($y$) velocity is
zero. The solution of the wake problem in this \emph{plasma}-frame is
of precisely the form we have considered above, except that the object
is moving with $y$-speed $-v_{\parallel\infty}$. Because we are
approximating the object as foreshortened in the scaled coordinates so
that its edge radius of curvature is negligible, \emph{the fact that
  it is moving is irrelevant} to the solution. When the $x$-axis is
interpreted as the time $c_st$ since the particular vertical slice of
plasma passed the object, the graphs we have plotted previously apply
\emph{without alteration}. If, instead, one were to plot a snap-shot
of the 2-dimensional spatial variation at a particular instant of
time, however, the $xy$ plane would be \emph{sheared} by the
$-v_{\parallel\infty}$ motion. This shearing is purely geometrical. It
amounts to the replacement of the $y$ spatial coordinate with $y'=y-
v_{\parallel\infty}t=y- (v_{\parallel\infty}/c_s)x$. Although this
shearing may be large (because of high Mach number) in the coordinates
we have been using, it does not affect the equations for the ion or
electron dynamics.

Therefore, the solutions we have obtained above apply directly to
cases with finite external parallel ion drift, provided that they are
interpreted as being the solutions as a function of \emph{time, in the plasma
frame of reference}.

The question then arises as to what the external electron velocity
distribution actually is, in this plasma reference frame. Is it
symmetric or not?  Since $v_{\parallel\infty}$ is typically $\sim10
c_s$ for the moon in the solar wind, the frame's velocity is a
significant fraction, $\sim 0.2$, of the electron thermal
velocity. If, then, the electron parallel distribution were a
stationary Maxwellian in the rest frame of the moon, it would be
substantially shifted in the moving frame and the effects of section
\ref{drift} would immediately apply. However, it can be shown
from considerations of magnetic field gradient that the
total electric current density in the solar wind must be far less than would
be implied by a relative velocity of electrons and ions of
$0.2v_e$. Therefore, in fact the mean electron and ion speeds are
very nearly equal in the external wind. That means the electron
distribution is \emph{unshifted} in the parallel-moving reference
frame, and the effects discussed in section \ref{drift} do not arise
from parallel drift (though they might arise from higher-order
electron distribution asymmetry such as skewness).

\section{Summary}

Two mechanisms by which unstable parallel electron velocity
distributions can arise in a magnetized plasma wake have been
explored. First (section \ref{drift}) substantial asymmetry of the
external velocity distribution, for example an overall parallel drift,
can be turned into instability by the wake's potential
structure. However, limits on electric current density in the solar
wind near the moon (for example) prevent average electron drift alone
from being large enough to generate major instability. Second (section
\ref{nonconserv}) the non-conservation of electron parallel energy in
the perpendicular drift also gives rise to unstable distribution
minima near the marginal electron velocity that only just traverses
the potential energy hill of the wake. The electron distributions have
been calculated for collisionless orbits, and the turbulence energy
density to which they would give rise has been evaluated
quasi-linearly. The instability is found to be everywhere in the wake
quite significant, and fairly strong near the wake axis. If
artificially large electron mass is used, as is frequently the case in
simulations that treat both electrons and ions by PIC techniques, then
this instability effect is greatly enhanced; so their results will not be in
quantitative agreement with nature.

Hybrid PIC simulations which proceed to the opposite extreme ---
infinitesimal electron mass
(e.g.\ \cite{Kallio2005,Travnicek2005,Wiehle2011})--- obviously omit
the parallel electron instability completely; but since they make no
pretence of treating the details of the electron dynamics, they will
perhaps be less likely to be misleading.  There are of course many
other ion and anisotropy instability mechanisms that will perturb the
electrons. The present work establishes the approximate Langmuir
turbulence level arising from the electron parallel distribution, with
which these other mechanisms will compete.

\section*{Acknowledgements}

I am grateful to Jasper Halekas and Stuart Bale for fascinating
discussions concerning space-craft measurements of the solar wind. 
Work supported in part by NSF/DOE Grant DE-FG02-06ER54982.

\bibliography{wake,Vacuum_Expansion,Lunar_Wake,Mine}

\ifagu
\end{article}
\fi

\end{document}